\documentclass[12pt,a4paper,english]{article}
\usepackage{titling}
\usepackage[affil-it]{authblk}
\usepackage{longtable}
\usepackage{lscape}
\usepackage{graphicx}
\usepackage{epsfig}
\usepackage{dsfont}
\usepackage{amsfonts}
\usepackage{amsmath}
\usepackage{amsthm}
\usepackage{amssymb}
\usepackage{bbold}
\usepackage[english]{babel}
\usepackage[utf8]{inputenc}
\usepackage[left=0.8in,right=0.8in,top=1.0in,bottom=1.2in]{geometry}
\usepackage{calc}
\usepackage{cancel}
\usepackage{float}
\usepackage{slashed}
\usepackage{mathrsfs}
\usepackage{pdfpages}
\usepackage{tabu}
\usepackage{simplewick}
\usepackage[hidelinks]{hyperref}
\usepackage{bm}
\usepackage{multirow}
\usepackage{hhline}
\usepackage{pifont}
\usepackage{float}
\usepackage[caption = false]{subfig}
\usepackage{xcolor}
\usepackage[normalem]{ulem}
\usepackage{cite}
\usepackage[compat=1.1.0]{tikz-feynman}
\tikzfeynmanset{warn luatex=false}
\usepackage{xcolor}
\usepackage{makecell}

\usepackage{datetime}
\usepackage{booktabs}

\numberwithin{equation}{section}
\numberwithin{figure}{section}
\numberwithin{table}{section}

\makeatletter
\g@addto@macro\bfseries{\boldmath}
\makeatother

\allowdisplaybreaks

\newcommand{\be}{\begin{equation}}
\newcommand{\ee}{\end{equation}}
\newcommand{\bea}{\begin{eqnarray}}
\newcommand{\eea}{\end{eqnarray}}
\newcommand{\ba}{\begin{align}}
\newcommand{\ea} {\end{align}}

\def\eq#1{{Eq.~(\ref{#1})}}
\def\eqs#1#2{{Eqs.~(\ref{#1})--(\ref{#2})}}
\def\fig#1{{Fig.~\ref{#1}}}

\def\Table#1{{Table~\ref{#1}}}

\def\sec#1{{Sect.~\ref{#1}}}

\def\app#1{{Appendix~\ref{#1}}}

\definecolor{ForestGreen}{RGB}{34,139,34}

\allowdisplaybreaks

\newcommand{\published}[1]{%
\gdef\puB{#1}}
\newcommand{\puB}{}

\date{}

\begin{document}

\title{\textbf{$\tau_{B_{s}}/\tau_{B_{d}}$ and $\Delta\Gamma_{s}$ confront new physics in $b\to s\tau\tau$}}

\author[1]{Marzia Bordone\thanks{marzia.bordone@cern.ch}}
\author[2]{Mario Fern\'andez Navarro\thanks{M.F.Navarro@soton.ac.uk}}
\affil[1]{Theoretical Physics Department, CERN, 1211 Geneva 23, Switzerland}
\affil[2]{School of Physics \& Astronomy, University of Southampton, Southampton SO17 1BJ, UK}

\published{\flushright
CERN-TH-2023-138\vskip2cm}

\maketitle

\begin{abstract}
Several new physics scenarios that address anomalies in $B$-physics predict an enhancement of $b \rightarrow s \tau \tau$ with respect to its Standard Model prediction. Such scenarios necessarily imply modifications of the lifetime ratio $\tau_{B_{s}}/\tau_{B_{d}}$ and the lifetime difference $\Delta\Gamma_{s}$. In this work, we explore indirect bounds provided by these observables over new physics scenarios. We also estimate future projections, showing that future experimental and theoretical improvements on both $\tau_{B_{s}}/\tau_{B_{d}}$ and $\Delta\Gamma_{s}$ have the potential to provide bounds competitive with those directly extracted from $b\rightarrow s \tau \tau$ transitions. After performing a model-independent analysis, we apply our results to the particular case of leptoquark mediators proposed to address the $R_{D^{(*)}}$ anomalies.
\end{abstract}

\section{Introduction}
The characterisation of new physics scenarios affecting flavour-changing processes is a challenging task. In fact, high theoretical and experimental accuracy is needed to extract significant constraints. Furthermore, a large set of observables is required to extract the complete flavour structure of the new physics couplings, and the compatibility between several constraints needs to be addressed.
Interestingly, in recent years hints of lepton flavour universality violation in $b\to c\tau\bar\nu$ mediated processes have raised a lot of attention. In fact, the lepton flavour universality ratios $R_{H_c} = \mathcal{B}(H_b\to H_c\tau\bar\nu)/\mathcal{B}(H_b\to H_c \mu\bar\nu)$, with $H_{b(c)}$ hadrons with a $b(c)$ quark, show deviations with respect to their respective Standard Model predictions. In particular, the two ratios $R_{D}$ and $R_{D^*}$ drive the discrepancy, at the level of 3.2 standard deviations \cite{HFLAV:2022pwe}. It is important to notice that, due to these deviations, a lot of progress has been achieved concerning Standard Model predictions, culminating with the impressive results from Lattice QCD in predicting $B\to D^{*}$ hadronic form factors \cite{FermilabLattice:2021cdg,Harrison:2023dzh,Aoki:2023qpa}. Despite the recent progress, these results show tensions among each other and with experimental data and hence require further investigation. For this work, we use the averaged predictions in \cite{HFLAV:2022pwe}. 
Other observables beyond $R_{D^{(*)}}$ have been measured, namely $R_{J/\psi}$ \cite{LHCb:2017vlu} and $R_{\Lambda_b}$ \cite{LHCb:2022piu}, which, however, are affected by large experimental errors. They are consistent with the current deviations in $R_{D^{(*)}}$, but they are not yet precise enough to shed light on this interesting puzzle.

In light of this, many Beyond the Standard Model scenarios have been hypothesized, with the introduction of new heavy states. Most of these scenarios, in order to accommodate the size of the discrepancy in $b\to c\tau\bar\nu$ decays, predict sizeable effects in $b\to s\tau\tau$ transitions, hence yielding a strong correlation between these two partonic processes. One of the most favourable ways of testing this correlation is using the bounds on $\mathcal{B}(B_s\to\tau\tau)$. However, with the increasing theoretical and experimental precision, it is natural to wonder whether it could be tested elsewhere. One of the possibilities is then looking at the lifetime difference of the $B_s$ system, or the lifetime ratio with respect to the $B_d$ meson. Both these observables are indeed modified by new physics in $b\to s\tau\tau$, and it can be studied if their foreseen precision allows extracting more information than with data on $\mathcal{B}(B_s\to\tau\tau)$. This is exactly the scope of this work. In \sec{sec:2} we briefly revise the theoretical framework and current status of the observables of interest. In \sec{sec:3}, we perform a model-independent analysis, highlighting correlations between the different observables contributing. In \sec{sec:4}, we work out the results for some explicit models. In \sec{sec:5}, we conclude.

\section{Setup and observables}
\label{sec:2}

The starting point of our analysis is to introduce New Physics (NP) in $b\to c\tau\bar\nu$ transitions. At the high scale $\mu=\Lambda$, where $SU(2)_L\times U(1)_Y$ invariance is restored, we have the following effective Lagrangian:
 \begin{equation}
\mathcal{L}^\mathrm{eff}=\mathcal{L}_\mathrm{SM}-\frac{1}{\Lambda^{2}}\sum_{i}\tilde{C}_{i}(\mu)\mathcal{Q}_{i}\,,
\label{eq:SMEFT}
\end{equation}
where the relevant operators, defined according to the SMEFT basis in \cite{Grzadkowski:2010es}, are listed in \app{app:A}. 
At the low scale $\mu=m_{b}$, $SU(2)_{L}\times U(1)_Y$ gauge invariance is broken and the effective Lagrangian reads: 
\begin{equation}
\mathcal{L}_{\mathrm{eff}}=-\frac{4G_{F}}{\sqrt{2}}\sum_{i}C_{i}(\mu)\mathcal{O}_{i}\label{eq:Lagrangian}
\end{equation}
where the operators $\mathcal{O}_i$ are in the LEFT basis \cite{Jenkins:2017dyc}. At the tree-level, only semileptonic operators are generated, and they are listed in \eqs{eq:basis1}{eq:basisfinal}.
The tree-level matching between the Wilson coefficients of the aforementioned
operators and the ones in the SMEFT is given in \app{app:A}, and the running between the scale $\mu=\Lambda\sim1\,\mathrm{TeV}$ and
$\mu=m_{b}$ is evaluated using DsixTools 2.1 \cite{Fuentes-Martin:2020zaz}. Non-zero contributions to four-quarks operators are induced at the loop level from semileptonic operators. The complete set of them is in \eqs{eq:4quarkWET1}{eq:4quarkWET3}.

In the next subsection, we
revise the mixing formalism of the $B_{s}$ meson system in presence of NP operators induced by the semileptonic ones, and discuss current status and prospects.

\subsection{\texorpdfstring{$\Delta\Gamma_{s}$}{Lifetime difference} and \texorpdfstring{$\tau_{B_{s}}/\tau_{B_{d}}$}{Lifetime ratio} beyond the Standard Model}
The absorptive off-diagonal element of the time evolution of the neutral $B_s$ meson system, $\Gamma_{12}^s$, and the dispersive one $M_{12}^s$, are closely related in the Standard Model. They define lifetimes,  mass differences and CP asymmetry as, respectively,
\begin{equation}
\Delta\Gamma_{s}=2|\Gamma_{12}^{s}|\cos\phi_{s}\,, \qquad \Delta M_s = |M_{12}^s|\,, \qquad a_{sl}^{s} = \mathrm{Im}\left(\frac{\Gamma_{12}^{s}}{M_{12}^s}\right)\,,
\end{equation}
with the mixing phase $\phi_{s}=\arg(-M_{12}^{s}/\Gamma_{12}^{s})$.
Corrections to these equations amount to $\mathcal{O}(1/8\left|\Gamma_{12}^{s}/M_{12}^{s}\right|^{2}\sin^{2}\phi_{s})\sim\mathcal{O}(10^{-11})$ in the SM, and are negligible with the current level of precision.

\begin{figure}
\begin{centering}
\begin{tikzpicture}
  \begin{feynman}
		\vertex (i1) {\(s\)};
		\vertex [above right=28mm of i1] (a);
		\vertex [above left=24mm of a] (i2) {\(b\)};
		\vertex [right=24mm of a] (b);
		\vertex [below right=24mm of b] (f1) {\(b\)};
		\vertex [above right=24mm of b] (f2) {\(s\)};
    \diagram* {
      (i1) -- [anti fermion] (a) -- [anti fermion] (i2) ,
      (a) -- [fermion, half left, edge label'=\(\tau\), near end] (b) -- [fermion, half left, edge label'=\(\tau\), near end] (a) ,
      (f1) -- [fermion] (b) -- [fermion] (f2) ,
    };

    \coordinate (midpoint) at ($(a)!0.5!(b)$);
    \draw [dashed, red] ($(midpoint) + (0, 3)$) -- ($(midpoint) + (0, -3)$);
  \end{feynman}
\end{tikzpicture}
\par\end{centering}
\caption{Diagram that contributes to $\Gamma^{s}_{12}$ via a double insertion of $(\bar{s}b)(\bar{\tau}\tau)$ operators. The tau loop in the diagram is closed and the cut (red dashed
line) indicates that only the imaginary part of the graph is taken.\label{fig:discontinuity_graph}}
\end{figure}
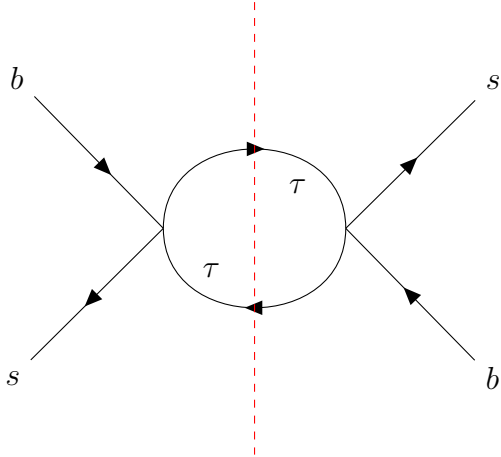

In case of NP couplings, the possible contributions to $\Delta M_s$ and $a_{sl}^s$ are closely linked to the UV completion of the various models. Hence, we postpone the discussion to the specific models in \sec{sec:4}. Instead, for $\Gamma_{12}^s$, the NP contributions are always finite, and are related to the discontinuity of the diagram in \fig{fig:discontinuity_graph}.
We follow the approach in Ref.~\cite{Bobeth:2011st}, where NP $(\bar{s}b)(\bar{\tau}\tau)$ effective operators are inserted. For each possible NP operator, we obtain:
\begin{flalign}
 & \left(\Gamma_{12}^{s}\right)_{S,AB}=3\mathcal{N}_{\Gamma_{12}^{s}}x_{\tau}\beta_{\tau}\left\langle Q_{S}^{B}\right\rangle \left([C_{ed}^{S,AB}]^{3323}\right)^{2}\,,\label{eq:GammaSAB}\\
 & \left(\Gamma_{12}^{s}\right)_{V,LA}=\mathcal{N}_{\Gamma_{12}^{s}}\beta_{\tau}\left[\left(1-x_{\tau}\right)\left\langle Q_{V}^{L}\right\rangle +\left(1+2x_{\tau}\right)\left\langle Q_{S}^{R}\right\rangle \right]\left([C_{ed}^{V,LA}]^{3323}\right)^{2}\,,\label{eq:GammaVAB}\\
 & \left(\Gamma_{12}^{s}\right)_{T,AA}=-12\mathcal{N}_{\Gamma_{12}^{s}}x_{\tau}\beta_{\tau}\left[4\left\langle Q_{S}^{A}\right\rangle +8\left\langle \widetilde{Q}_{S}^{A}\right\rangle \right]\left([C_{ed}^{T,AA}]^{3323}\right)^{2}\,\label{eq:GammaTAB}.
\end{flalign}
where $A=L,R$, $B=L,R$ and we have introduced $\beta_{\tau}=\sqrt{1-4x_{\tau}}$, $x_{\tau}=m_{\tau}^{2}/m_{b}^{2}$
and $\mathcal{N}_{\Gamma_{12}^{s}}=-G_{F}^{2}m_{b}^{2}/(6\pi M_{B_{s}})$.  The corresponding expression for $\left(\Gamma_{12}^{s}\right)_{V,RA}$ can be obtained by exchanging the labels $L$ and $R$ everywhere. Note that each NP contribution is obtained by insertion of twice the same operator, since the interference between different operators is zero. Therefore, in the presence of more than one operator, the total NP contribution to $\Gamma_{12}^s$ can be obtained by summing the contribution of each individual operator. While reviewing the calculation, we found a sign difference with respect to Eq.(E.3) in Ref.~\cite{Bobeth:2011st}. However, despite of this discrepancy, the final result in \eqs{eq:GammaSAB}{eq:GammaTAB} agrees with Eq.~(60) in Ref.~\cite{Bobeth:2011st}.
We define the matrix elements of the possible NP operators in the $B_s$ field as
\begin{flalign*}
  & \left\langle Q_{V}^{A}\right\rangle =\left\langle \bar{B}_{s}|\right.(\bar{s}\gamma_{\mu}P_{A}b)(\bar{s}\gamma^{\mu}P_{A}b)\left.|B_{s}\right\rangle =\frac{2}{3}f_{B_{s}}^{2}M_{B_{s}}^{2}B_{B_{s}}^{(1)}\,,\\
  & \left\langle Q_{S}^{A}\right\rangle =\left\langle \bar{B}_{s}|\right.(\bar{s}P_{A}b)(\bar{s}P_{A}b)\left.|B_{s}\right\rangle =-\frac{5}{12}\left(\frac{M_{B_{s}}}{m_{b}(\mu)+m_{s}(\mu)}\right)^{2}f_{B_{s}}^{2}M_{B_{s}}^{2}B_{B_{s}}^{(2)}\,,\\
  & \left\langle \tilde{Q}_{S}^{A}\right\rangle =\left\langle \bar{B}_{s}|\right.(\bar{s}_{\alpha}P_{A}b_{\beta})(\bar{s}_{\beta}P_{A}b_{\alpha})\left.|B_{s}\right\rangle =\frac{1}{12}\left(\frac{M_{B_{s}}}{m_{b}(\mu)+m_{s}(\mu)}\right)^{2}f_{B_{s}}^{2}M_{B_{s}}^{2}B_{B_{s}}^{(3)}\,.
\end{flalign*}
where $\alpha$ and $\beta$ are color indices. The numerical input for the expressions above, including values for the bag parameters $B_{B_{s}}^{(i)}$,
is given in Appendix \ref{sec:Lattice_input}.

The ratio $\Delta\Gamma_{s}/\Delta\Gamma_{s}^{\mathrm{SM}}$ is related to $\Gamma_{12}^{s,\mathrm{NP}}$ through the following general expression:
\begin{equation}
\frac{\Delta\Gamma_{s}}{\Delta\Gamma_{s}^{\mathrm{SM}}}=\left|1+\frac{\Gamma_{12}^{s,\mathrm{NP}}}{\Gamma_{12}^{s,\mathrm{SM}}}\right|\frac{\cos\left[\phi_{s}^{\mathrm{SM}}+\arg(1+\frac{\Gamma_{12}^{s,\mathrm{NP}}}{\Gamma_{12}^{s,\mathrm{SM}}})\right]}{\cos\phi_{s}^{\mathrm{SM}}}\,.
\end{equation}
If we assume real NP Wilson coefficients, i.e.~$\Gamma_{12}^{s,\mathrm{NP}}$
is real, and neglect $\phi_{s}^{\mathrm{SM}}$ along with the small phase
of $\Gamma_{12}^{s,\mathrm{SM}}$, then we arrive to the common formula
found in the literature \cite{Bobeth:2011st,Bobeth:2014rda}

\begin{equation}
\frac{\Delta\Gamma_{s}}{\Delta\Gamma_{s}^{\mathrm{SM}}}=1+\frac{\Gamma_{12}^{s,\mathrm{NP}}}{|\Gamma_{12}^{s,\mathrm{SM}}|}\,.
\end{equation}
Using the current SM prediction for $\Delta\Gamma_s$ \cite{Gerlach:2022hoj}, and the HFLAV average \cite{HFLAV:2022pwe} we obtain for the ratio of the decay width difference
\begin{equation}
\frac{\Delta\Gamma_{s}}{\Delta\Gamma_{s}^{\mathrm{SM}}}=1.11\pm0.26\,,
\end{equation}
which defines the space allowed for new physics. Similarly, the future projection is given by \cite{LenzFCC:2022}
\begin{equation}
\left(\frac{\Delta\Gamma_{s}}{\Delta\Gamma_{s}^{\mathrm{SM}}}\right)_{2035}=1.06\pm0.06 \,.
\label{eq:projection_DeltaGamma}
\end{equation}
With this input, we can extract model independent bounds on the size of NP contributions. By using the current data we find
\begin{equation}
\Gamma_{12}^{s,\mathrm{NP}}<0.022\,\mathrm{ps}^{-1}\,,\qquad(95\%\,\mathrm{CL})
\end{equation}
 and considering the 2035 numerical projection for $\Delta\Gamma_{s}/\Delta\Gamma_{s}^{\mathrm{SM}}$ included in Eq.~\eqref{eq:projection_DeltaGamma}, we obtain the projected bound
\begin{equation}
\left(\Gamma_{12}^{s,\mathrm{NP}}\right)_{2035}<0.0029\,\mathrm{ps}^{-1}\,.\qquad(95\%\,\mathrm{CL})
\end{equation}

\begin{table}[t]
\begin{centering}
\begin{tabular}{|c|c|c|c|c|}
\cline{2-5} \cline{3-5} \cline{4-5} \cline{5-5} 
\multicolumn{1}{c|}{} &  & \multicolumn{3}{c|}{$\mathcal{B}(B_{s}\rightarrow\tau\tau)_{\mathrm{NP}}$}\tabularnewline
\hline 
Assumptions  & Input  & $\tau_{B_{s}}/\tau_{B_{d}}$  & LHCb ($50\,\mathrm{fb}^{-1}$)  & LHCb (300 $\mathrm{fb}^{-1}$)\tabularnewline
\hline 
\multirow{3}{*}{H1} & \multirow{2}{*}{$\left(\frac{\tau_{B_{s}}}{\tau_{B_{d}}}\right)_{\mathrm{SM}}=1.020(5)$} & \multirow{4}{*}{$1.7(6)\cdot10^{-2}$} &  & \multicolumn{1}{c|}{}\tabularnewline
 &  &  &  & \tabularnewline
\cline{2-2} 
 & $\left(\frac{\tau_{B_{s}}}{\tau_{B_{d}}}\right)_{\mathrm{exp}}=1.001(1)$  &  & \multirow{3}{*}{$<1.3\cdot10^{-3}$} & \multirow{3}{*}{$<5\cdot10^{-4}$}\tabularnewline
\cline{1-3} \cline{2-3} \cline{3-3} 
\multirow{3}{*}{H2} & \multirow{2}{*}{$\left(\frac{\tau_{B_{s}}}{\tau_{B_{d}}}\right)_{\mathrm{SM}}=1.001(1)$} & \multirow{4}{*}{$<2.6\cdot10^{-3}$} &  & \tabularnewline
 &  &  &  & \tabularnewline
\cline{2-2} 
 & $\left(\frac{\tau_{B_{s}}}{\tau_{B_{d}}}\right)_{\mathrm{exp}}=1.001(1)$  &  &  & \tabularnewline
\hline 
\end{tabular}
\par\end{centering}
\caption{Projected bounds at 95\% CL for $\mathcal{B}(B_{s}\rightarrow\tau\tau)_{\mathrm{NP}}$
obtained from the lifetime ratio $\tau_{B_{s}}/\tau_{B_{d}}$ are
confronted against the projected bounds from LHCb \cite{LHCb:2018roe}.
For the projections in the lifetime ratio, we assume that the uncertainties
will reduce to 1 per mille in the experiment.
We display different results under two different hypothesis for the SM prediction: that the central value will remain as the current
one and the $SU(3)_{F}$ breaking parameters
could be measured to a 10\% precision (H1), and that the central value will shift to match the experiment and there is no $SU(3)_{F}$ breaking up to the per mille level (H2).}
\label{tab:Lifetime_Projected} 
\end{table}

The NP effects in $\bar{s}b\bar{\tau}\tau$ couplings affect also the lifetime ratio of $B_{s}$ and $B_{d}$ mesons. If we assume no NP effects in the $B_d$ lifetime, we have:
\begin{equation}
\frac{\tau_{B_{s}}}{\tau_{B_{d}}}=\left(\frac{\tau_{B_{s}}}{\tau_{B_{d}}}\right)_{\mathrm{SM}}\left(1+\frac{\Gamma(B_{s}\rightarrow\tau\tau)_{\mathrm{NP}}}{\Gamma(B_{s})_{\mathrm{SM}}}\right)^{-1}\,,
\label{eq:LifetimeRatio_NP}
\end{equation}
where we define $\Gamma(B_{s}\rightarrow\tau\tau)_\mathrm{NP}=\Gamma(B_{s}\rightarrow\tau\tau)_\mathrm{total}-\Gamma(B_{s}\rightarrow\tau\tau)_\mathrm{SM}$, which encodes the NP contribution to the partial decay width. The expression for $\Gamma(B_{s}\rightarrow\tau\tau)_\mathrm{total}$ can be extracted from \eq{eq:Bs_tautau}.
The SM prediction for the lifetime ratio can be found in Ref.~\cite{Lenz:2022rbq}, and it depends on non-perturbative parameters in the Heavy Quark Expansion as well as the size of $SU(3)_f$ breaking between the $B_s$ and the $B_d$ system. We employ the central values and errors for the expectation values of the next-to-leading power matrix element in the $B_d$ field from \cite{Bordone:2021oof}. Concerning the size of $SU(3)_f$ breaking, estimates using Heavy-Quark Effective Theory relations \cite{Bordone:2022qez,Lenz:2022rbq} and preliminary Lattice QCD estimations \cite{Gambino:2017vkx,Gambino:2017dfa} are affected by large errors. To be very conservative, we use the central values from \cite{Bordone:2022qez} and assign $100\%$ errors. With this, we obtain:
\begin{align}
\left(\frac{\tau_{B_{s}}}{\tau_{B_{d}}}\right)_{\mathrm{SM}}=1.02\pm 0.02 \,, &  & 
\Gamma(B_{s})_{\mathrm{SM}}=0.597_{-0.069}^{+0.106}\,\mathrm{ps^{-1}}\,,\label{eq:LifetimeRatio_SM}
\end{align}
that is compared with the current experimental HFLAV average \cite{HFLAV:2022pwe},
\begin{equation}
\left(\frac{\tau_{B_{s}}}{\tau_{B_{d}}}\right)_{\mathrm{HFLAV\,2022}}=1.001\pm0.004\,.
\end{equation}
At the current status, we find good agreement between the SM predictions and the lifetime average, albeit with large uncertainties due to the unknown $SU(3)_f$ breaking.

We note that in the literature, it has been discussed the impact of using the values from a  different set of non perturbative parameters in the lifetime ratio \cite{Bernlochner:2022ucr,Lenz:2022rbq}, which yield to a large shift. However, it has to be noticed that the values for these parameters change a lot depending on whether higher dimensional operators are considered or not, hinting to non-trivial correlations. This is not observed in \cite{Bordone:2021oof}, that we adopt as our reference.
We can now extract an indirect limit over NP contributions to $\mathcal{B}(B_s\to\tau\tau)$ from the lifetime ratio. Using \eq{eq:LifetimeRatio_NP}, we obtain
\begin{equation}
\mathcal{B}(B_{s}\rightarrow\tau\tau)_{\mathrm{NP}}
<5.5 \times 10^{-2}\,,
\end{equation}
at the 95\% CL, which has to be compared with the direct bound from LHCb \cite{LHCb:2017myy}, namely $\mathcal{B}(B_{s}\rightarrow\tau\tau)_{\mathrm{NP}}
<6.8 \times 10^{-3}$.  Currently, the direct bound over $\mathcal{B}(B_{s}\rightarrow\tau\tau)_{\mathrm{NP}}$ obtained by the LHCb
collaboration is 40\% better than the indirect bound obtained from the lifetime ratio.

We then repeat this comparison with the projected sensitivities. The results are shown in \Table{tab:Lifetime_Projected}. For the experimental measurement, we explore the possibility that the error will reduce to 1 per mille. For the SM prediction, we explore two hypotheses corresponding to either no change in the central value or a substantial reduction of it, towards a strong indication of small $SU(3)_f$ breaking. In hypothesis H1, we assume that the $SU(3)_f$ breaking parameters could be measured to a 10\% precision, as possible in the foreseeable future using Lattice QCD, but retaining the current central values, while in H2 we impose no $SU(3)_f$ breaking up to the per mill level. The LHCb collaboration provides two expected upper bounds for $\mathcal{B}(B_s\to\tau\tau)$: a first projection is based on a luminosity of 50$\mathrm{fb^{-1}}$, which in contrast to the expectations in \cite{LHCb:2018roe} will be reached only after 2032. The second upper bound from the LHCb collaboration is based on an expected luminosity of 300$\mathrm{fb^{-1}}$, which with respect of the expectations in \cite{LHCb:2018roe}, will be reached only after 2041. This shows that improved measurements and predictions of the lifetime ratios have the potential of improving the current bound on $\mathcal{B}(B_s\to\tau\tau)$, while waiting for LHCb to collect the necessary statistics to obtain even more stringent bounds.
 This motivates extra efforts both from  the theoretical and experimental communities to investigate $\tau_{B_{s}}/\tau_{B_{d}}$ as a potential channel to constrain NP effects. 


\section{Model independent analysis}
\label{sec:3}
\begin{figure}[p]
\begin{centering}
\subfloat[\label{fig:parameter_space_Vectors}]{\includegraphics[scale=0.4]{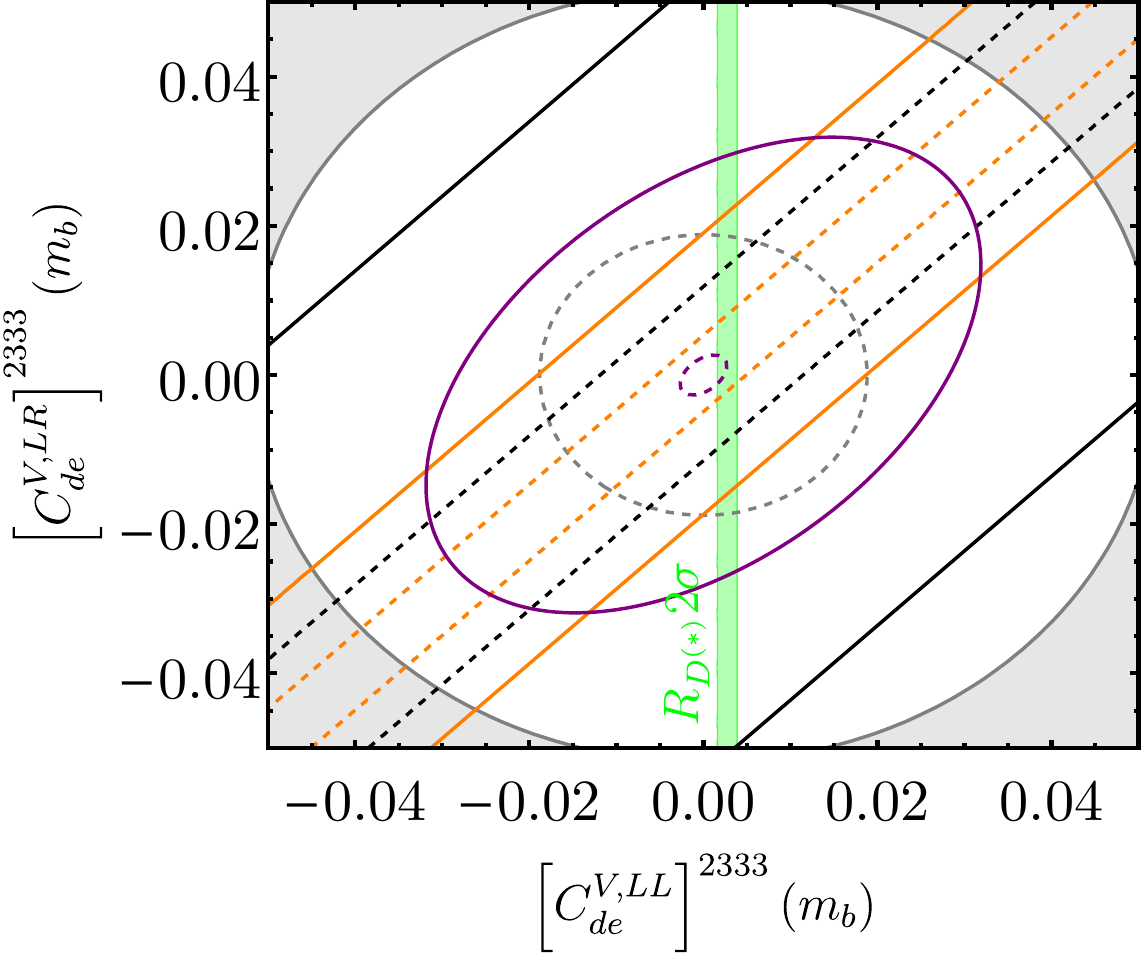}

}$\qquad$\subfloat[\label{fig:parameter_space_Scalars}]{\includegraphics[scale=0.43]{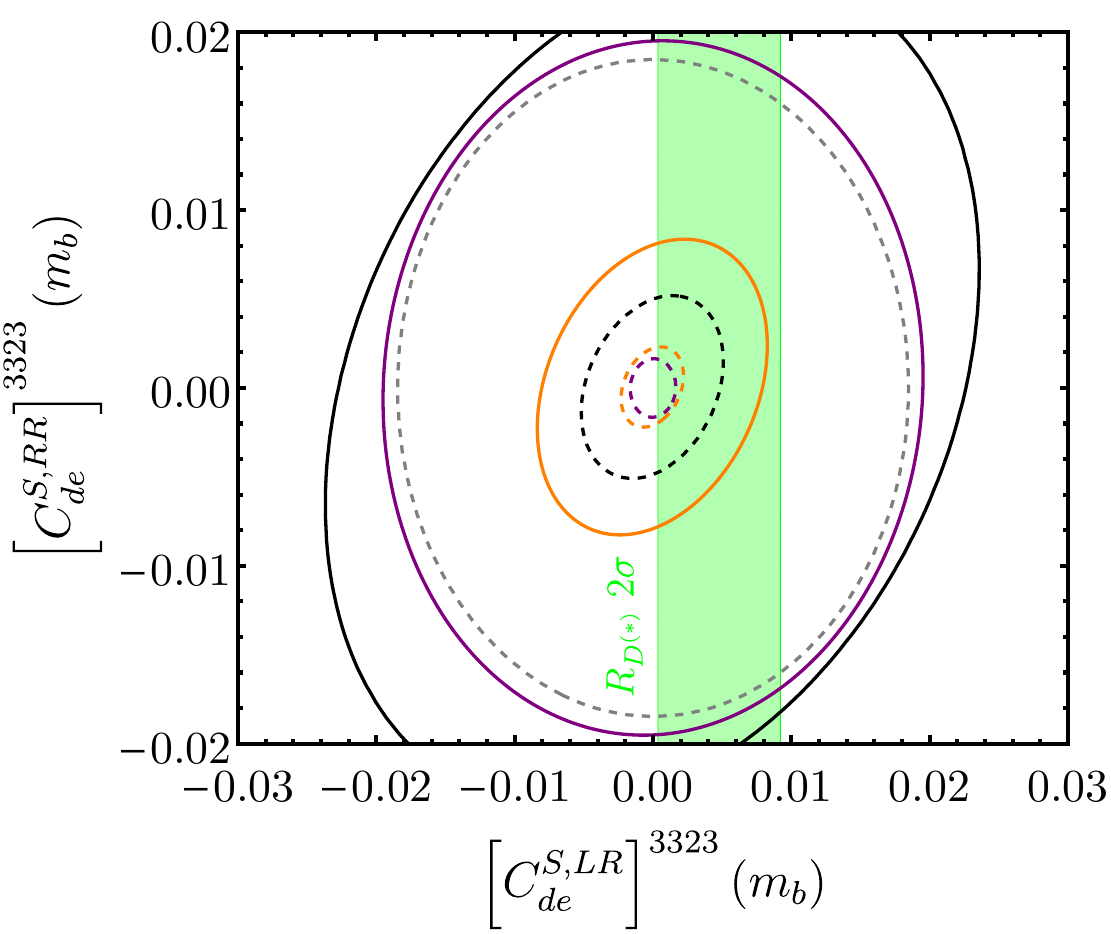}

}
\par\end{centering}
\begin{centering}
\subfloat[\label{fig:C10_C10prime}]{\includegraphics[scale=0.38]{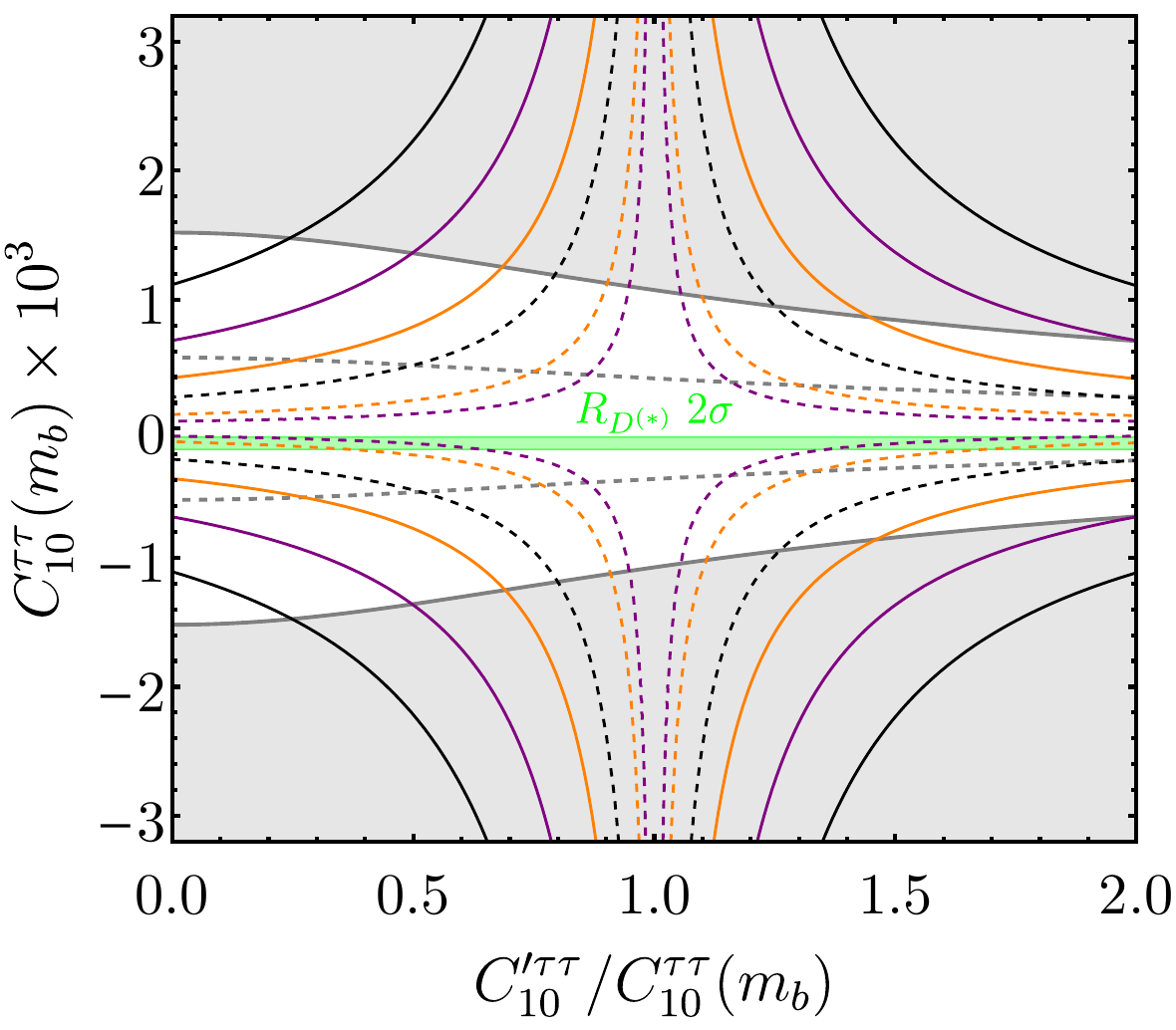}

}$\qquad$\subfloat[\label{fig:CS_CSprime}]{\includegraphics[scale=0.4]{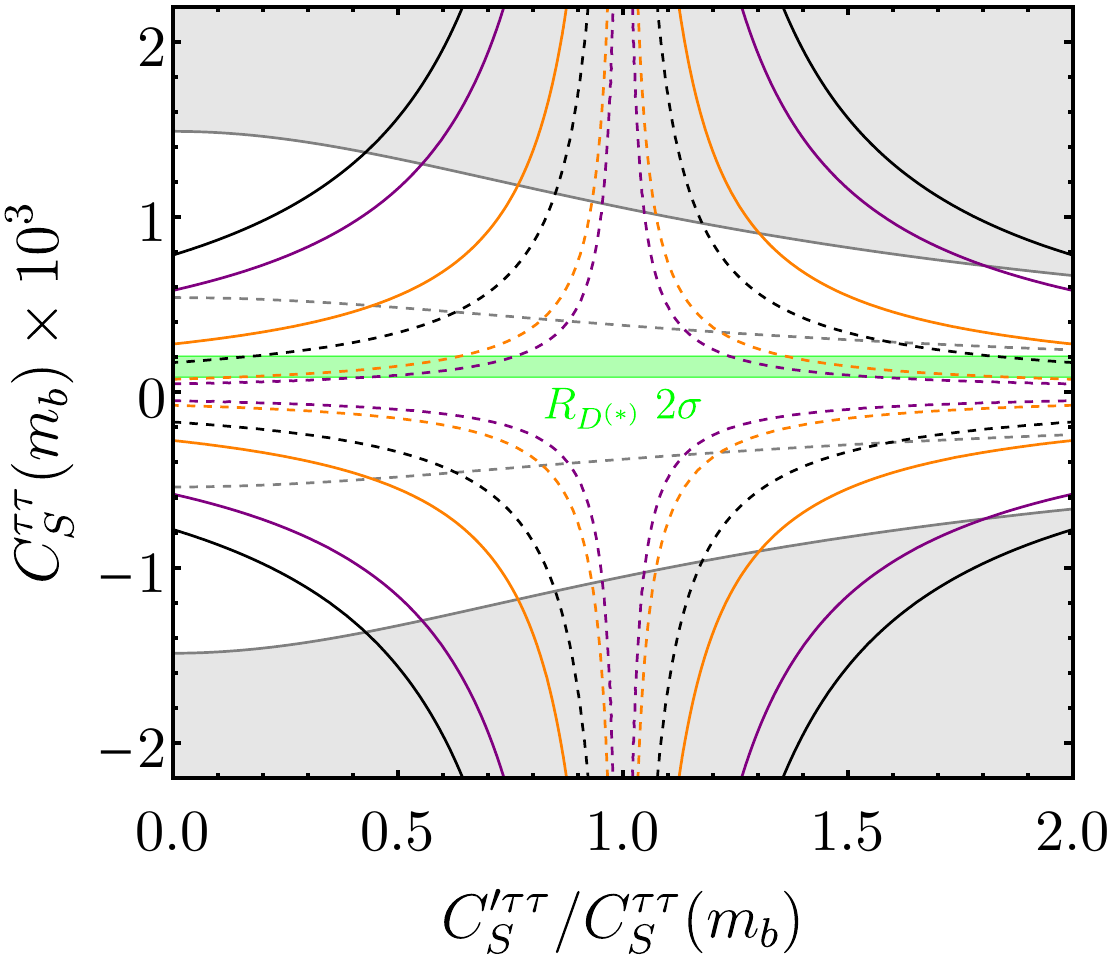}

}
\par\end{centering}
\caption{(\textbf{\textit{Top}}) Parameter space of vector (left) and scalar
(right) Wilson coefficients (see the main text). (\textbf{\textit{Bottom}})
Absolute size of $C_{10}^{\tau\tau}(m_{b})$ vs the ratio $C_{10}^{\tau\tau}(m_{b})/C'^{\tau\tau}_{10}(m_{b})$
(left) and $C_{S}^{\tau\tau}(m_{b})$ vs the ratio $C_{S}^{\tau\tau}(m_{b})/C'^{\tau\tau}_{S}(m_{b})$ (right).
The plots in the bottom allow to study the parameter space in the scenario where
both $C_{10(S)}^{\tau\tau}$ and $C'^{\tau\tau}_{10(S)}$ have similar
size and sign. In all panels, orange contours represent the direct bounds over $\mathcal{B}(B_{s}\rightarrow\tau\tau)$,
while the black contour represent the indirect bounds over $\mathcal{B}(B_{s}\rightarrow\tau\tau)$ obtained from
$\tau_{B_{s}}/\tau_{B_{d}}$, the purple contours represent the direct
bounds over $\mathcal{B}(B^{+}\rightarrow K^{+}\tau\tau)$, the grey contours represent the direct bounds over $\Delta\Gamma_{s}/\Delta\Gamma_{s}^{\mathrm{SM}}$, and the green region is preferred by $R_{D^{(*)}}$ at $2\sigma$. Solid (dashed) contours denote current (projected) 95\% CL exclusions. For $\mathcal{B}(B_{s}\rightarrow\tau\tau)$, we show the projected direct bound by LHCb with 300 $\mathrm{fb}^{-1}$.}
\end{figure}
In this section, we employ the previous results to obtain bounds on
the NP Wilson coefficients in a model independent way. We start by assuming that only the vector operators
$[C_{ed}^{V,LL}]^{3323}$ and $[C_{de}^{V,LR}]^{3323}$ are non-zero. These operators are interesting because they contribute to both $C_{9}^{\tau\tau}$
and $C_{10}^{\tau\tau}$ via \eq{eq:C9C10_CedV}, which potentially receive bounds from our set of $b\rightarrow s \tau\tau$ observables. Moreover, we assume that $SU(2)_{L}$ invariance at high energies implies the connection,
\begin{equation}
[C_{ed}^{V,LL}]^{3323}\approx[C_{\nu edu}^{V,LL}]^{3332}\,,
\end{equation}
which leads to correlations with $R_{D^{(*)}}$. The parameter space of
$[C_{ed}^{V,LL}]^{3323}$ and $[C_{de}^{V,LR}]^{3323}$ is depicted
in Fig.~\ref{fig:parameter_space_Vectors}. Currently, the most competitive bounds over the parameter space are given by the direct bounds on $\mathcal{B}(B_{s}\rightarrow\tau\tau)$ and $\mathcal{B}(B^{+}\rightarrow K^{+}\tau\tau)$, followed by the indirect bounds obtained from the lifetime ratio $\tau_{B_{s}}/\tau_{B_{d}}$ and from the lifetime difference $\Delta\Gamma_{s}$. More interesting, however, is the future projected picture of the parameter space. Eventually, the strongest bounds are expected to come from updates by LHCb and Belle II, which will nevertheless require the collection of substantial integrated luminosity \cite{LHCb:2018roe,Belle-II:2018jsg}. Following from our discussion of future projections in \sec{sec:2}, it is not unreasonable that the lifetime ratio sets the strongest constraints over $C_{10}^{\tau\tau}$-related scenarios for a certain period in the near future. Similarly, $\Delta\Gamma_{s}$ has the potential to set competitive constraints over scenarios involving both $C_{9}^{\tau\tau}$ and $C_{10}^{\tau\tau}$ until the next update of $\mathcal{B}(B^{+}\rightarrow K^{+}\tau\tau)$ by Belle II.

In the particular NP scenarios where both primed and unprimed operators have similar size and sign, the NP contributions to $\mathcal{B}(B_{s}\rightarrow\tau\tau)$ and $\mathcal{B}(B^{+}\rightarrow K^{+}\tau\tau)$ cancel, and the most competitive constraint becomes that of $\Delta\Gamma_{s}$. This is illustrated in \fig{fig:C10_C10prime}, where only $C_{10}^{\tau\tau}$ and $C'^{\tau\tau}_{10}$ are non-zero. We obtain that for $C'^{\tau\tau}_{10}/C^{\tau\tau}_{10}\approx1.4$ the bound from $\Delta\Gamma_{s}$ is already the most competitive. This shows that potentially, better measurements of $\Delta\Gamma_s$ can help distinguish new physics scenarios characterised by similar NP couplings for left-handed and right-handed quarks, which naturally provide accidental cancellations in $\mathcal{B}(B_s\to\tau\tau)$.

In the following, we explore the parameter space of the scalar operators $[C_{ed}^{S,LL}]^{3323}$ and $[C_{ed}^{S,LR}]^{3323}$, which contribute to $C_{S}^{\tau\tau}$
and $C_{P}^{\tau\tau}$ via  Eq.~\eqref{eq:CSCP_CedS}. We assume that $SU(2)_{L}$ invariance at high energies implies the connection,
\begin{equation}
[C_{ed}^{S,LR}]^{3323}\approx([C_{\nu edu}^{S,RL}]^{3332})^{*}\,.
\end{equation}
We illustrate the different bounds over the parameter space in  Fig.~\ref{fig:parameter_space_Scalars}. Direct bounds over $\mathcal{B}(B_{s}\rightarrow\tau\tau)_{\mathrm{NP}}$ and $\mathcal{B}(B^{+}\rightarrow K^{+}\tau\tau)$ provide the strongest constraints over the parameter space when considering both current and projected bounds, but the lifetime ratio has the potential to provide very competitive bounds in the near future. Just like in the case of vector operators, in the particular NP scenarios where both primed and unprimed scalar operators have similar size and sign, the NP contributions to $\mathcal{B}(B_{s}\rightarrow\tau\tau)$ and $\mathcal{B}(B^{+}\rightarrow K^{+}\tau\tau)$ cancel, and the most competitive constraint becomes that of $\Delta\Gamma_{s}$. In this way, in Fig.~\ref{fig:CS_CSprime} we explore the scenario where only $C_{S}^{\tau\tau}$ and $C'^{\tau\tau}_{S}$ are non-zero. We obtain that for $C'^{\tau\tau}_{S}/C^{\tau\tau}_{S}\approx1.2$ the bound from $\Delta\Gamma_{s}$ is already the most competitive.

Finally, we note that, as already stated in Ref.~\cite{Capdevila:2017iqn}, additional constraints will arise from future measurements of $B\to K^*\tau^+\tau^-$ and $B_s\to\phi\tau^+\tau^-$. In fact, being these decays a pseudovector to vector transition, they probe different combinations of Wilson coefficients, and provide stricter bounds on $C_9^{\tau\tau}$, hypothesising that the same potential sensitivity as for $B^+\to K^+\tau^+\tau^-$ will be reached. Future projections taking into account the upcoming and current upgrade of the LHCb and Belle II experiments, can shed light on the actual constraining power of these modes.

\section{New physics models}
\label{sec:4}
There is a rich literature about new mediators beyond the SM that
have been proposed to explain the $R_{D^{(*)}}$ anomalies (see e.g.~\cite{Dorsner:2016wpm,Angelescu:2021lln}). In particular,
leptoquarks constitute some of the most promising NP candidates, as
they can naturally provide the required semileptonic operators while
avoiding tree-level contributions to $\Delta M_{s}$. Nevertheless,
some of them predict an enhancement of $b\rightarrow s\tau\tau$
observables much above their SM prediction, therefore potentially
undergoing the constraints from $\tau_{B_{s}}/\tau_{B_{d}}$ and $\Delta\Gamma_{s}$
derived in the previous sections. Furthermore, it has been shown that
particular leptoquarks can also provide a one-loop and lepton universal
contribution to the operator $C_{9}$, which has been shown to greatly
improve the global fit to $b\rightarrow s\ell\ell$ observables provided that the NP
effect is roughly one fourth of the SM $C_{9}$ \cite{Alguero:2023jeh,SinghChundawat:2022ldm}.
We denote such universal contribution as $C_{9}^{U}$, which specific
leptoquarks can provide via RGE mixing from $C_{9}^{\tau\tau}$ and $C_{10}^{\tau\tau}$
\cite{Capdevila:2017iqn,Crivellin:2018yvo}.

We start by discussing scalar leptoquarks. Our conclusions can be summarised as in the following:
\begin{itemize}
\item The $S_{1}\sim(\mathbf{\bar{3}},\mathbf{1},1/3)$ leptoquark generates tree-level contributions to $R_{D^{(*)}}$, while $b\rightarrow s\tau\tau$ is only induced at loop level. Therefore, the correlation between these two modes is less strong and we checked explicitly that the loop induced effects in $b\to s\tau\tau$ are negligible for all the projected bounds considered
in this paper.
\item The $S_{3}\sim(\mathbf{\bar{3}},\mathbf{3},1/3)$ leptoquark induces both $R_{D^{(*)}}$ and $b\rightarrow s\tau\tau$ at the tree-level. However, leading constraints from $B\rightarrow K^{(*)}\nu\nu$ and
$\Delta M_{s}$ render impossible to fully address the anomaly in $R_{D^{(*)}}$.
\item The leptoquark $R_{2}\sim(\mathbf{3},\mathbf{2},7/6)$ provides uncorrelated contributions to $R_{D^{(*)}}$ and $b\to s\tau\tau$. The latter generates  $C_{9}^{U}<0$ via RGE mixing,
therefore receiving constraints from $\tau_{B_{s}}/\tau_{B_{d}}$
and $\Delta\Gamma_{s}$. However, we find that the leading constraint to this scenario comes from $\Delta M_{s}$ at 1-loop, that requires $\left|C_{9}^{U}\right|<0.3$ in agreement
with previous analyses \cite{Crivellin:2022mff}. Although a particular cancellation
mechanism could alleviate the constraint from $\Delta M_{s}$, this would require further model building beyond the scope of this work.
Ultimately, even if $\Delta M_{s}$ is alleviated, the projected constraint
from $\tau_{B_{s}}/\tau_{B_{d}}$ could only rule out $\left|C_{9}^{U}\right|>2$,
unable to reach the region $C_{9}^{U}\approx-1$ preferred by global fits \cite{Alguero:2023jeh}.

\end{itemize}
Vector leptoquarks are instead more promising. In the next
subsections, we will study the ones that address
$R_{D^{(*)}}$ while providing chirally enhanced contributions to
$\mathcal{B}(B_{s}\rightarrow\tau\tau)$.

\subsection{The vector leptoquark \texorpdfstring{$U_{1}\sim(\mathbf{3},\mathbf{1},2/3)$}{U1}}

\begin{figure}[t]

\centering
\includegraphics[scale=0.54]{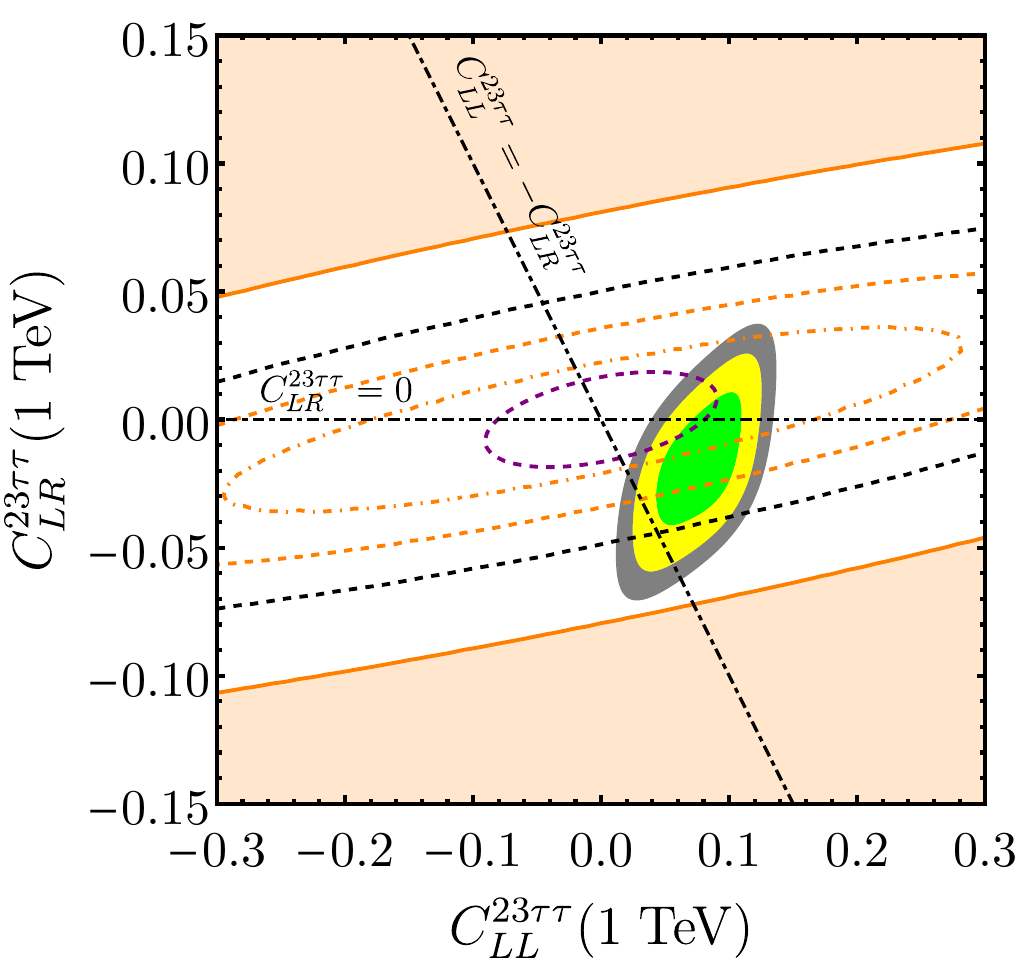}

\caption{Parameter space of Wilson coefficients motivated by the $U_{1}$ vector leptoquark explanation of $R_{D^{(*)}}$ (see main text). The green, yellow and grey regions represent the $1\sigma$, $2\sigma$ and $3\sigma$ regions preferred by $R_{D^{(*)}}$, respectively. Orange contours represent the direct bounds from $\mathcal{B}(B_{s}\rightarrow\tau\tau)$,
while the black contour represents the indirect bound obtained from
$\tau_{B_{s}}/\tau_{B_{d}}$ and the purple contour represents the direct bounds from $\mathcal{B}(B^{+}\rightarrow K^{+}\tau\tau)$. Solid (dashed) contours denote current (projected) 95\% CL exclusions, except for the two projections for $\mathcal{B}(B_{s}\rightarrow\tau\tau)$ by LHCb: 50 $\mathrm{fb}^{-1}$ (orange dashed) and 300 $\mathrm{fb}^{-1}$ (orange dash-dotted). Dash-dotted black lines represent two interesting benchmark scenarios motivated in the main text.} \label{Fig:U1}
\end{figure}

The $U_{1}\sim(\mathbf{3},\mathbf{1},2/3)$ vector leptoquark is a well
motivated mediator to explain the $R_{D}$ and $R_{D^{*}}$ anomalies
\cite{Cornella:2019hct,Cornella:2021sby,Greljo:2018tuh,Calibbi:2017qbu,Blanke:2018sro,DiLuzio:2017vat,DiLuzio:2018zxy,Aebischer:2022oqe}.
A well motivated embedding for gauge $U_{1}$ leptoquarks is the Pati-Salam group \cite{Pati:1974yy}, that provides a natural connection with quark-lepton
unification.
Moreover, explanations of the $R_{D^{(*)}}$ anomalies via exchange of  the $U_{1}$ vector leptoquark had been shown to be naturally connected with the origin of flavour
hierarchies and the flavour structure of the SM \cite{Bordone:2017bld,Bordone:2018nbg,Allwicher:2020esa,Fuentes-Martin:2022xnb,Davighi:2022bqf,King:2021jeo,FernandezNavarro:2022gst}.
Remarkably, the contributions of the $U_{1}$ vector leptoquark to $R_{D^{(*)}}$ are correlated
to an enhancement of $b\rightarrow s\tau\tau$.
At an effective scale $\Lambda$ higher than the electroweak scale, the $U_1$ interactions are well described in the context of the SMEFT as:
\begin{equation}
\mathcal{L}_{\mathrm{SMEFT}}^{U_{1}}\supset-\frac{1}{\Lambda^{2}}\left[\frac{C_{LL}^{ij\alpha\beta}}{2}\left(\mathcal{Q}_{\ell q}^{(1)}+\mathcal{Q}_{\ell q}^{(3)}\right)^{ij\alpha\beta}-\left(2C_{LR}^{ij\alpha\beta}\left(\mathcal{Q}_{\ell edq}^{\dagger}\right)^{ij\alpha\beta}+\mathrm{h.c.}\right)\right]\,.
\end{equation}
The matching between the relevant LEFT and SMEFT Wilson coefficients
reads: 
\begin{align}
\left[C_{\nu edu}^{V,LL}\right]^{3332*}(m_{b})= & \eta_{V}^{\tau\nu}C_{LL}^{23\tau\tau}(\Lambda)\frac{v^{2}}{2V_{cb}\Lambda^{2}} & \left[C_{\nu edu}^{S,RL}\right]^{3332*}(m_{b})= & -\eta_{S}^{\tau\nu}2C_{LR}^{23\tau\tau}(\Lambda)\frac{v^{2}}{2V_{cb}\Lambda^{2}}\,,\label{eq:U1_ops1}\\
\left[C_{ed}^{V,LL}\right]^{3323}(m_{b})= & \eta_{V}^{\tau\tau}C_{LL}^{23\tau\tau}(\Lambda)\frac{v^{2}}{2\Lambda^{2}}\,, & \left[C_{ed}^{S,LR}\right]^{3323}(m_{b})= & -\eta_{S}^{\tau\tau}2C_{LR}^{23\tau\tau}(\Lambda)\frac{v^{2}}{2\Lambda^{2}}\,,\label{eq:U1_ops3}
\end{align}
where the factors $\eta_{i}^{\tau\tau}$ and $\eta_{i}^{\tau\nu}$
encode the running from the high scale $\Lambda=1\,\mathrm{TeV}$
and are evaluated with DsixTools \cite{Celis:2017hod}, obtaining
$\eta_{V}^{\tau\tau}\simeq0.96$, $\eta_{S}^{\tau\tau}\simeq1.57$
$\eta_{V}^{\tau\nu}\simeq1.03$ and $\eta_{S}^{\tau\nu}\simeq1.64$.
The operators in Eq.~(\ref{eq:U1_ops3}) match into $C_{9}^{\tau\tau}=-C_{10}^{\tau\tau}$
and $C_{S}^{\tau\tau}=-C_{P}^{\tau\tau}$ via Eq.~(\ref{eq:C9C10_CedV})
and Eq.~(\ref{eq:CSCP_CedS}), respectively. The presence of the scalar operator
$\left[C_{ed}^{S,LR}\right]^{3323}$, which ultimately provides $C_{S}^{\tau\tau}=-C_{P}^{\tau\tau}$,
delivers a chirally enhanced contribution to $\mathcal{B}(B_{s}\rightarrow\tau\tau)$
connected to the size of $C_{LR}^{23\tau\tau}$. If $C_{LR}^{23\tau\tau}=0$,
then $\mathcal{B}(B_{s}\rightarrow\tau\tau)$ is still substantially
enhanced by the presence of $C_{9}^{\tau\tau}=-C_{10}^{\tau\tau}$,
but chiral enhancement is lost.

In Fig.~\ref{Fig:U1} we explore the parameter space of SMEFT Wilson
coefficients in the model, highlighting two particularly motivated
benchmark scenarios. The case $C_{LL}^{23\tau\tau}=-C_{LR}^{23\tau\tau}$
is a good benchmark for 4321 models featuring TeV scale third family quark-lepton
unification \cite{Bordone:2017bld,Cornella:2019hct,Cornella:2021sby,Greljo:2018tuh,Fuentes-Martin:2022xnb,Davighi:2022bqf},
while the case $C_{LR}^{23\tau\tau}=0$ is a good benchmark for the
flavour universal 4321 model \cite{DiLuzio:2017vat,DiLuzio:2018zxy,King:2021jeo,FernandezNavarro:2022gst}. Given that $U_{1}$ is a vector leptoquark, the leading contribution to $\Delta M_{s}$ arising at 1-loop depends on the specific UV completion. For the well-motivated case of 4321 models, the contribution to $\Delta M_{s}$ is dominated by a vector-like lepton running in the loop, and the most stringent constraints can be avoided as long as the mass of the vector-like lepton is around or below the TeV scale \cite{DiLuzio:2018zxy,Fuentes-Martin:2020hvc,Cornella:2021sby,FernandezNavarro:2022gst}. In this manner, the model is able to address $R_{D^{(*)}}$ and the enhancement of $\mathcal{B}(B_{s}\rightarrow\tau\tau)$ becomes a key prediction of the model. 

Due to chiral enhancement, $\mathcal{B}(B_{s}\rightarrow\tau\tau)$
is particularly sensitive to scenarios with large $\left|C_{LR}^{23\tau\tau}\right|$,
but current direct bounds from LHCb cannot yet test the preferred
region by the benchmark case $C_{LL}^{23\tau\tau}=-C_{LR}^{23\tau\tau}$.
Remarkably, in the near future we expect the indirect bound from
the lifetime ratio $\tau_{B_{s}}/\tau_{B_{d}}$ to constrain a region
of the parameter space preferred by $C_{LL}^{23\tau\tau}=-C_{LR}^{23\tau\tau}$,
while the parameter space preferred by $C_{LR}^{23\tau\tau}=0$ is
expected to remain unconstrained. In the longer term, updated direct
measurements of $\mathcal{B}(B_{s}\rightarrow\tau\tau)$ and $\mathcal{B}(B^{+}\rightarrow K^{+}\tau\tau)$
have the potential to test most of the parameter space compatible
with $R_{D}$ and $R_{D^{*}}$.

As a final remark, using the results in \cite{DiLuzio:2018zxy,Fuentes-Martin:2020hvc,Cornella:2021sby,FernandezNavarro:2022gst}, and our aforementioned results for $\Gamma_{12}^{s}$, we estimated the NP impact on $a_{sl}^s$. However, due to due absence of a NP phase in the relevant couplings, the NP contribution to $a_{sl}^s$ is dominated by the phase of $V_{ts}^*$ multiplied by small couplings. Hence, in this scenario we find no visible effect in $a_{sl}^s$. We notice that for complex right-handed couplings, this would not be the case, and would be worth studying it in detail if the misalignment between $R_D$ and $R_{D^{*}}$ changes significantly with new measurements.

\subsection{The vector leptoquark \texorpdfstring{$V_{2}\sim(\mathbf{\bar{3}},\mathbf{2},5/6)$}{V2}}

The vector leptoquark $V_{2}\sim(\mathbf{\bar{3}},\mathbf{2},5/6)$ arises in the context of grand unified theories (GUTs) based on the $SU(5)$ gauge group.
In a recent work \cite{Iguro:2023prq}, it has been pointed out that a TeV scale $V_{2}$ vector leptoquark could explain the deviations in $R_{D^{(*)}}$ via the
scalar operator $(\bar{c}_{L}b_{R})(\bar{\tau}_{R}\nu_{L\tau})$,
arising from the SMEFT operator $[\mathcal{Q}^\dagger_{\ell edq}]^{\tau\tau23}$.
Notice that this operator is also predicted by the $U_{1}$ vector
leptoquark in models featuring third family quark-lepton unification
at the TeV scale, and it has the interesting feature of correlating
the enhancement of $R_{D^{(*)}}$ with a chiral enhancement of $\mathcal{B}(B_{s}\rightarrow\tau\tau)$.
For the purpose of this work, we shall work with a simplified phenomenological
Lagrangian, requiring only the minimal couplings needed to address
the $R_{D^{(*)}}$ anomalies. In this manner, the di-quark coupling
that would lead to a rapid proton decay is also absent. The relevant
interaction terms read 

\begin{equation}
\mathcal{L}_{V_{2}}\supset\beta_{i\alpha}^{dL}\left(\overline{d^{C}}_{i}\gamma_{\mu}L_{\alpha}^{b}\right)\epsilon_{ab}V_{2}^{\mu,a}+\beta_{i\alpha}^{Qe}\left(\overline{Q^{C,a}}_{i}\gamma_{\mu}e_{\alpha}\right)\epsilon_{ab}V_{2}^{\mu,b}+\mathrm{h.c.}
\label{eq:V2_Lagrangian}
\end{equation}
where 
\begin{equation}
\beta_{i\alpha}^{dL}=\left(\begin{array}{ccc}
0 & 0 & 0\\
0 & 0 & 0\\
0 & 0 & \beta_{3\tau}^{dL}
\end{array}\right)\,,\quad\beta_{i\alpha}^{Qe}=\left(\begin{array}{ccc}
0 & 0 & 0\\
0 & 0 & \beta_{2\tau}^{Qe}\\
0 & 0 & 0
\end{array}\right)\,.
\label{eq:V2_couplings}
\end{equation}

As usual for leptoquarks, $V_{2}$ does not contribute to  $\Delta M_{s}$ at tree level. Using DsixTools 2.1~\cite{Fuentes-Martin:2020zaz}, we have studied the 1-loop RGE mixing of $[\mathcal{Q}^{\dagger}_{\ell edq}]^{\tau\tau23}$ into low energy operators that could potentially contribute to $\Delta M_{s}$ and $a_{sl}^s$. We find all operators to receive vanishing contributions, with the exception of $[L^{V,LL}_{dd}]^{2323}$ that receives a negligible contribution at the level 0.0001\% of the SM contribution. We stress that, in a UV complete model, we expect further states to be generated when breaking the $SU(5)$ group to the SM. They can potentially generate further contributions to $\Delta M_{s}$, that depend on the specific breaking chain. A study of the UV completion for the $V_2$ vector leptoquark is, however, beyond the scope of this work, but will be required for a comprehensive analysis of loop-induced constraints on this vector state.
\begin{figure}[t]

\centering
\includegraphics[scale=0.54]{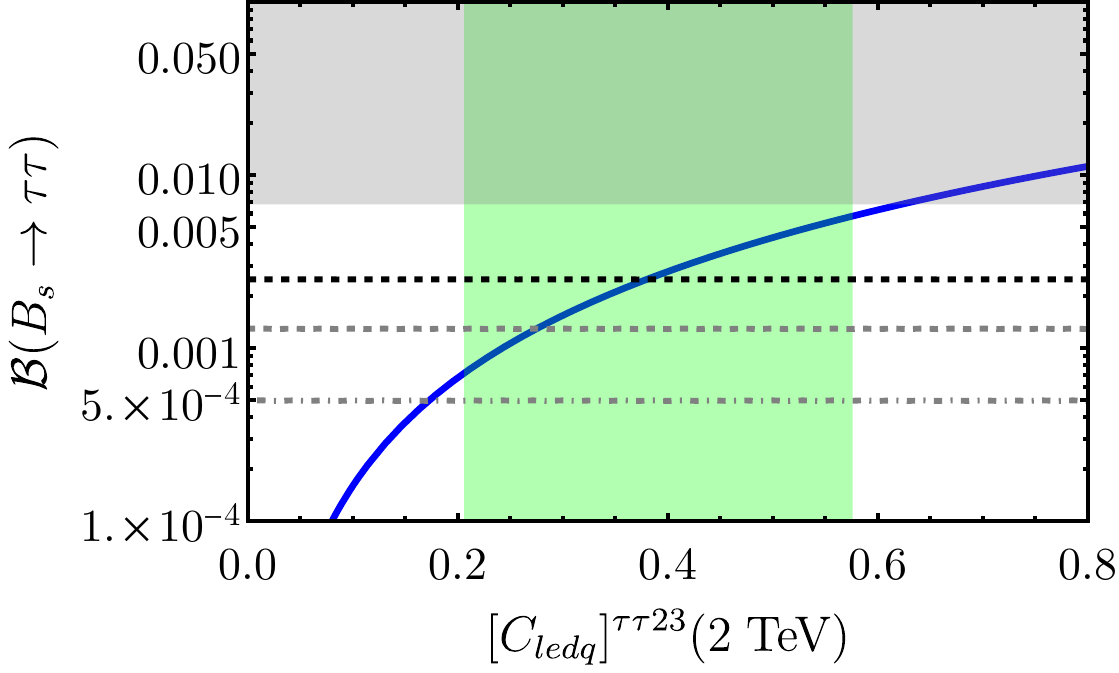}

\caption{$V_{2}$ model prediction for $\mathcal{B}(B_{s}\rightarrow\tau\tau)$ (blue line) as a function of the SMEFT Wilson coefficient $[C_{\ell edq}]^{\tau\tau32}$. The current 95\% CL excluded region by LHCb is grey shaded. The dashed black line depicts the projected 95\% CL indirect bound obtained from $\tau_{B_{s}}/\tau_{B_{d}}$. The gray dashed and dashed-dotted lines represent the projected 95\% CL direct bounds by LHCb for $50\:\mathrm{fb^{-1}}$ and $300\:\mathrm{fb^{-1}}$, respectively \label{Fig:V2}}
\end{figure}

Remarkably, if we work in the basis of mass eigenstates $Q_{i}=(V_{ij}u_{L}^{j},d_{L}^{i})$
where the CKM mixing originates from the up sector, then a contribution
to $\mathcal{B}(B_{u}\rightarrow\tau\nu)$ severely constrains the
model. Nevertheless, this contribution can be easily suppressed by
introducing $\beta_{1\tau}^{Qe}$ and enforcing some mild cancellation
with $\beta_{2\tau}^{Qe}$ \cite{Iguro:2023prq}. After integrating out $V_{2}$
and matching to the SMEFT, we obtain the following operator at tree-level
\begin{equation}
\mathcal{L}_{\mathrm{SMEFT}}^{V_{2}}=\frac{[C_{\ell edq}]^{\tau\tau32}}{M_{V_{2}}^{2}}[\mathcal{Q^{\dagger}}_{\ell edq}]^{\tau\tau23}=\frac{2\beta_{3\tau}^{dL}(\beta_{2\tau}^{Qe})^{*}}{M_{V_{2}}^{2}}[\mathcal{Q^{\dagger}}_{\ell edq}]^{\tau\tau23}\,,
\end{equation}
which at low energies matches into,
\begin{equation}
\left[C_{\nu edu}^{S,RL}\right]^{3332*}(m_{b})=-\eta_{S}^{\tau\nu}[C_{\ell edq}]^{\tau\tau32}(M_{V_{2}})\frac{v^{2}}{2V_{cb}M_{V_{2}}^{2}}\,,
\end{equation}
\begin{equation}
\left[C_{ed}^{S,LR}\right]^{3323}(m_{b})=-\eta_{S}^{\tau\tau}[C_{\ell edq}]^{\tau\tau32}(M_{V_{2}})\frac{v^{2}}{2M_{V_{2}}^{2}}\,.
\end{equation}
Notice that this is the same operator predicted by the $U_{1}$ leptoquark
in models featuring third family quark-lepton unification, as discussed
in the previous section. As such, it provides chiral enhancement of
$\mathcal{B}(B_{s}\rightarrow\tau\tau)$ via $C_{S}^{\tau\tau}=-C_{P}^{\tau\tau}$,
obtained from $\left[C_{ed}^{S,LR}\right]^{3323}$ when applying
Eq.~(\ref{eq:C9C10_CedV}).

The scalar operator $\left[C_{\nu edu}^{S,RL}\right]^{3332*}$ provides
a large contribution to $R_{D}$, able to fit the current experimental central value, while the contribution to $R_{D^{*}}$ accommodates only marginally the current tension (see Eqs.~\eqref{eq:RD} and \eqref{eq:RD*}). In Fig.~\ref{Fig:V2}, we show the model
prediction for $\mathcal{B}(B_{s}\rightarrow\tau\tau)$ as a function
of $[C_{\ell edq}]^{\tau\tau32}\:(2\,\mathrm{TeV})$. We can see that the current
direct bound by LHCb is unable to constrain the region preferred at $1\sigma$ by $R_{D}$. Nevertheless, in the near
future we expect the indirect bound from the lifetime ratio $\tau_{B_{s}}/\tau_{B_{d}}$
to provide a leading constraint over the model. Being more specific,
$\tau_{B_{s}}/\tau_{B_{d}}$ will constrain the model from explaining
the central values of $R_{D}$ (or larger). In the much longer term,
LHCb has the potential to fully test the model with
$300\:\mathrm{fb}^{-1}$ of integrated luminosity.

\section{Conclusions}
\label{sec:5}
Several new physics scenarios proposed to address anomalies in $B$-physics naturally predict an enhancement of $b\rightarrow s \tau \tau$. In this work, we have explored the impact of new physics in the $b\rightarrow s \tau \tau$ channel over the lifetime ratio $\tau_{B_{s}}/\tau_{B_{d}}$ and the lifetime difference $\Delta\Gamma_{s}$. First of all, via a model-independent analysis, we assessed the constraining power of the lifetime ratio and lifetime difference over NP in $b\rightarrow s \tau \tau$.
We conclude that such observables provide indirect bounds over new physics scenarios, which, however, are not currently competitive with the existing direct experimental bounds. Nevertheless, we have estimated future projections and concluded that both $\tau_{B_{s}}/\tau_{B_{d}}$ and $\Delta\Gamma_{s}$ can provide competitive bounds before the LHCb and Belle II experiments reach the large integrated luminosities required to improve their direct bounds on $b\rightarrow s \tau \tau$ transitions. By looking at the different NP operators, we find that the lifetime ratio can potentially constrain scenarios where $\mathcal{B}(B_s\to\tau\tau)$ is enhanced. On the other hand, the lifetime difference is very interesting in scenarios where $\mathcal{B}(B_s\to\tau\tau)$ is not modified by NP couplings, and also to constrain scenarios with similar NP couplings for left-handed and right-handed quarks. 
We also introduce simplified models that can address $R_{D^{(*)}}$ without generating tree-level contributions to the neutral meson mass differences. Two scenarios are particularly interesting, namely the ones of the vector leptoquarks $U_{1}\sim(\mathbf{3},\mathbf{1},2/3)$ and $V_{2}\sim(\mathbf{\bar{3}},\mathbf{2},5/6)$. In fact, in these scenarios $\tau_{B_{s}}/\tau_{B_{d}}$ can provide competitive constraints in the near future, thanks to a chiral enhancement of $\mathcal{B}(B_{s}\rightarrow\tau\tau)$ provided by scalar low-energy operators. This work motivates efforts by both the theoretical and experimental communities to investigate $\tau_{B_{s}}/\tau_{B_{d}}$ and $\Delta\Gamma_{s}$ as potential channels to constrain NP effects.

\section*{Acknowledgements}
We thank Claudia Cornella and Gino Isidori for useful comments on this manuscript. We thank Miguel Escudero for pointing out a typo in \app{app:C}. The work of MFN is supported by the European Union’s Horizon 2020 Research and Innovation Programme under Marie Skłodowska-Curie grant agreement HIDDeN
European ITN project (H2020-MSCA-ITN-2019//860881-HIDDeN). MFN is grateful to the CERN Theory Division for the hospitality during parts of this work.

\appendix

\section{Operator basis}
\label{app:A}
In this Appendix, we list all the operators that we use in our analysis. In the SMEFT \cite{Grzadkowski:2010es} we have:
\begin{align}
[\mathcal{Q}_{lq}^{(1)}]^{\alpha\beta ij}= & \,(\bar{L}^{\alpha}\gamma^{\mu}L^{\beta})(\bar{Q}^{i}\gamma^{\mu}Q^{j}) & \,[\mathcal{Q}_{lq}^{(3)}]^{\alpha\beta ij}= & \,(\bar{L}^{\alpha}\gamma^{\mu}\sigma^{a}L^{\beta})(\bar{Q}^{i}\gamma^{\mu}\sigma^{a}Q^{j}) \label{eq:operators_SMEFT_1}\\{}
[\mathcal{Q}_{ed}]^{\alpha\beta ij}= & \,(\bar{e}_{R}^{\alpha}\gamma_{\mu}e_{R}^{\beta})(\bar{d}_{R}^{i}\gamma^{\mu}d_{R}^{j}) & \,[\mathcal{Q}_{ld}]^{\alpha\beta ij}= & \,(\bar{L}^{\alpha}\gamma^{\mu}L^{\beta})(\bar{d}_{R}^{i}\gamma^{\mu}d_{R}^{j})\label{eq:operators_SMEFT_2}\\{}
[\mathcal{Q}_{ledq}]^{\alpha\beta ij}= & \,(\bar{L}_{a}^{\alpha}e_{R}^{\beta})(\bar{d}_{R}^{i}Q_{a}^{j}) & [\mathcal{Q}_{qe}]^{\alpha\beta ij}= &\, (\bar{Q}^{\alpha}\gamma^{\mu}Q^{\beta})(\bar{e}_{R}^{i}\gamma^{\mu}e_{R}^{j})\,.\label{eq:operators_SMEFT_3}
\end{align}
 and we work in the basis: $Q^{i,T}=(V_{ij}^{*}u_{L}^{j},d_{L}^{i})$
and $L^{i,T}=(U_{ij}^{*}\nu_{j},e_{L}^{i})$, where $V$ and $U$ denote the CKM and PMNS matrices, respectively.\\
 In the LEFT \cite{Jenkins:2017dyc,Jenkins:2017jig}, we use the following FCNCs operators:
\begin{align}
[\mathcal{O}_{ed}^{V,LL}]^{\alpha\beta ij}= & \,(\bar{\ell}^{\alpha}\gamma_{\mu}P_{L}\ell^{\beta})(\bar{d}^{i}\gamma^{\mu}P_{L}d^{j}) & [\mathcal{O}_{ed}^{V,RR}]^{\alpha\beta ij}= & \,(\bar{\ell}^{\alpha}\gamma_{\mu}P_{R}\ell^{\beta})(\bar{d}^{i}\gamma^{\mu}P_{R}d^{j})\label{eq:basis1}\\{}
[\mathcal{O}_{ed}^{V,LR}]^{\alpha\beta ij}= & \,(\bar{\ell}^{\alpha}\gamma_{\mu}P_{L}\beta)(\bar{d}^{i}\gamma^{\mu}P_{R}d^{j}) & [\mathcal{O}_{ed}^{V,RL}]^{\alpha\beta ij}= & \,(\bar{\ell}^{\alpha}\gamma_{\mu}P_{R}\ell^{\beta})(\bar{d}^{i}\gamma^{\mu}P_{L}d^{j})\\{}
[\mathcal{O}_{ed}^{S,RR}]^{\alpha\beta ij}= & \,(\bar{\ell}^{\alpha}P_{R}\ell^{\beta})(\bar{d}^{i}P_{R}d^{j}) & [\mathcal{O}_{ed}^{S,RL}]^{\alpha\beta ij}= & \,(\bar{\ell}^{\alpha}P_{R}\ell^{\beta})(\bar{d}^{i}P_{L}d^{j})\\{}
[\mathcal{O}_{ed}^{S,LL}]^{\alpha\beta ij}= & \,(\bar{\ell}^{\alpha}P_{L}\ell^{\beta})(\bar{d}^{i}P_{L}d^{j}) & [\mathcal{O}_{ed}^{S,LR}]^{\alpha\beta ij}= & \,(\bar{\ell}^{\alpha}P_{L}\ell^{\beta})(\bar{d}^{i}P_{R}d^{j})\\{}
[\mathcal{O}_{ed}^{T,RR}]^{\alpha\beta ij}= & (\bar{\ell}^{\alpha}\sigma_{\mu\nu}P_{R}\ell^{\beta})(\bar{d}^{i}\sigma^{\mu\nu}P_{R}d^{j}) & \,[\mathcal{O}_{ed}^{T,LL}]^{\alpha\beta ij}= & \,(\bar{\ell}^{\alpha}\sigma_{\mu\nu}P_{L}\ell^{\beta})(\bar{d}^{i}\sigma^{\mu\nu}P_{L}d^{j})\,,\label{eq:basis7}
\end{align}
the charged current operators:
\begin{align}
[\mathcal{O}_{\nu edu}^{V,LL}]^{\alpha\beta ij}= & \,(\bar{\nu}^{\alpha}\gamma^{\mu}P_{L}\ell^{\beta})(\bar{d}^{i}\gamma_{\mu}P_{L}u^{j}) & [\mathcal{O}_{\nu edu}^{V,LR}]^{\alpha\beta ij}= & \,(\bar{\nu}\alpha\gamma^{\mu}P_{L}\ell^{\beta})(\bar{d}^{i}\gamma_{\mu}P_{R}u^{j})\\{}
[\mathcal{O}_{\nu edu}^{S,RR}]^{\alpha\beta ij}= & \,(\bar{\nu}^{\alpha}P_{R}\ell^{\beta})(\bar{d}^{i}P_{R}u^{j}) & [\mathcal{O}_{\nu edu}^{S,RL}]^{\alpha\beta ij}= & \,(\bar{\nu}^{\alpha}P_{R}\ell^{\beta})(\bar{d}^{i}P_{L}u^{j})\\{}
[\mathcal{O}_{\nu edu}^{T,RR}]^{\alpha\beta ij}= & \,(\bar{\nu}^{\alpha}\sigma^{\mu\nu}P_{R}\ell^{\beta})(\bar{d}^{i}\sigma^{\mu\nu}P_{R}u^{j}) \label{eq:basisfinal}
\end{align}
and the four-quark operators:
\begin{align}
[\mathcal{O}_{dd}^{V,LL}]^{ijkl}= & \,(\bar{d}_{L}^{i}\gamma_{\mu}d_{L}^{j})(\bar{d}_{L}^{k}\gamma^{\mu}d_{L}^{l}) & [\mathcal{O}_{dd}^{V,RR}]^{ijkl}= & \,(\bar{d}_{R}^{i}\gamma_{\mu}d_{R}^{j})(\bar{d}_{R}^{k}\gamma^{\mu}d_{R}^{l})\label{eq:4quarkWET1}\\{}
[\mathcal{O}_{dd}^{V1,LR}]^{ijkl}= & \,(\bar{d}_{L}^{i}\gamma_{\mu}d_{L}^{j})(\bar{d}_{R}^{k}\gamma^{\mu}d_{R}^{l}) & [\mathcal{O}_{dd}^{V8,RR}]^{ijkl}= & \,(\bar{d}_{L}^{i}\gamma_{\mu}T^{a}d_{L}^{j})(\bar{d}_{R}^{k}\gamma^{\mu}T^{a}d_{R}^{l})\\{}
[\mathcal{O}_{dd}^{S1,RR}]^{ijkl}= & \,(\bar{d}_{L}^{i}d_{R}^{j})(\bar{d}_{L}^{k}d_{R}^{l}) & [\mathcal{O}_{dd}^{S8,RR}]^{ijkl}= & \,(\bar{d}_{L}^{i}T^{a}d_{R}^{j})(\bar{d}_{L}^{k}T^{a}d_{R}^{l}) \label{eq:4quarkWET3}
\end{align}
\\
Following the normalization of Eqs.~\eqref{eq:SMEFT} and \eqref{eq:Lagrangian}, the tree-level matching for the operators relevant for $b\to s\tau\tau$ reads \cite{Jenkins:2017jig} 
\begin{align}
[C_{ed}^{V,LL}]^{3323}= & \,\frac{[\tilde{C}_{\ell q}^{(1)}]^{3323}v^{2}}{2\Lambda^{2}}+\frac{[\tilde{C}_{\ell q}^{(3)}]^{3323}v^{2}}{2\Lambda^{2}}
\,, & [C_{ed}^{V,RR}]^{3323}= & \,\frac{[\tilde{C}_{ed}]^{3323}v^{2}}{2\Lambda^{2}}\,,
\\
[C_{ed}^{V,LR}]^{3323}= & \,\frac{[\tilde{C}_{\ell d}]^{3323}v^{2}}{2\Lambda^{2}}\,,
& 
[C_{de}^{V,LR}]^{3323}= & \,\frac{[\tilde{C}_{qe}]^{3323}v^{2}}{2\Lambda^{2}}\,,
\\
[C_{ed}^{S,RR}]^{3323}= & \,0\,, & [C_{ed}^{S,RL}]^{3323}= & \,\frac{[\tilde{C}_{\ell edq}]^{3323}v^{2}}{2\Lambda^{2}}\,,\\{}
[C_{ed}^{S,LL}]^{3323}= & \,0\,, & [C_{ed}^{S,LR}]^{3323}= & \,\frac{[(\tilde{C}_{\ell edq})^{*}]^{3332}v^{2}}{2\Lambda^{2}}\,,\\{}
[C_{ed}^{T,RR}]^{3323}= &\, 0\,, & [\mathcal{O}_{ed}^{T,LL}]^{3323}= &\, 0\,.
\end{align}
For the operators relevant for $b\to c\tau\bar\nu$, following the normalization
of Eqs.~\eqref{eq:SMEFT} and \eqref{eq:btoctaunu_low} we have
\begin{align}
[C_{\nu edu}^{V,LL}]^{3332}= & \frac{2[\tilde{C}_{\ell q}^{(3)}]^{3332}v^{2}}{2V_{cb}\Lambda^{2}}\,, & [C_{\nu edu}^{V,LR}]^{3332}= & 0\,,\\{}
[\mathcal{C}_{\nu edu}^{S,RR}]^{\alpha\beta ij}= & \frac{[\tilde{C}_{\ell equ}^{(3)}]^{3332}v^{2}}{2V_{cb}\Lambda^{2}}\,, & [C_{\nu edu}^{S,RL}]^{3332}= & \frac{[\tilde{C}_{\ell edq}]^{3332}v^{2}}{2V_{cb}\Lambda^{2}}\,,\\{}
[C_{\nu edu}^{T,RR}]^{3332}= & \frac{[\tilde{C}_{\ell equ}^{(3)}]^{3332}v^{2}}{2V_{cb}\Lambda^{2}}\,.
\end{align}
where in all cases above we are assuming that NP do not modify the
$Z$ boson nor $W$ boson couplings to fermions.

\section{Lattice input}

\label{sec:Lattice_input}
The BSM bag parameters are taken from the averages in \cite{Dowdall:2019bea} and
displayed in Table \ref{tab:Bag_Parameters}. They refer to the operators: 

\begin{table}[t]
 
\centering{}%
\begin{tabular}{cccccc}
\toprule 
$i$ & 1 & 2 & 3 & 4 & 5\tabularnewline
\midrule
$B_{B_{s}}^{(i)}$ & 0.84(3) & 0.83(4) & 0.85(5) & 1.03(4) & 0.94(3)\tabularnewline
\bottomrule
\end{tabular}
\caption{Bag parameters $B_{B_{s}}^{(i)}$ in the $\overline{\mathrm{MS}}$
scheme evaluated at the scale $\mu=\overline{m}_{b}$. The values
displayed are the averages taken from \cite{Dowdall:2019bea}. \label{tab:Bag_Parameters}}
\vspace{0.1in}
\end{table}

\begin{equation}
\mathcal{O}_{1}^{s}=\left(\bar{b}\text{\ensuremath{\gamma_{\mu}}}P_{L}s\right)\left(\bar{b}\text{\ensuremath{\gamma^{\mu}}}P_{L}s\right)\,,
\end{equation}

\begin{equation}
\mathcal{O}_{2}^{s}=\left(\bar{b}P_{L}s\right)\left(\bar{b}P_{L}s\right)\,,\qquad\mathcal{O}_{3}^{s}=\left(\bar{b}_{\alpha}P_{L}s_{\beta}\right)\left(\bar{b}_{\beta}P_{L}s_{\alpha}\right)\,,
\end{equation}

\begin{equation}
\mathcal{O}_{4}^{s}=\left(\bar{b}P_{L}s\right)\left(\bar{b}P_{R}s\right)\,,\quad\mathcal{O}_{5}^{s}=\left(\bar{b}_{\alpha}P_{L}s_{\beta}\right)\left(\bar{b}_{\beta}P_{R}s_{\alpha}\right)\,.
\end{equation}
Their expectation values are given by:
\begin{equation}
\left\langle \mathcal{O}_{1}^{s}\right\rangle =c_{1}f_{B_{q}}^{2}M_{B_{s}}^{2}B_{B_{s}}^{(1)}(\mu)
\end{equation}

\begin{equation}
\left\langle \mathcal{O}_{i}^{s}\right\rangle =c_{i}\left(\frac{M_{B_{s}}}{m_{b}(\mu)+m_{s}(\mu)}\right)^{2}f_{B_{q}}^{2}M_{B_{s}}^{2}B_{B_{s}}^{(i)}(\mu)\,,\quad i=2,3
\end{equation}

\begin{equation}
\left\langle \mathcal{O}_{i}^{s}\right\rangle =c_{i}\left(\frac{M_{B_{s}}}{m_{b}(\mu)+m_{s}(\mu)}+d_{i}\right)^{2}f_{B_{q}}^{2}M_{B_{s}}^{2}B_{B_{s}}^{(i)}(\mu)\,,\quad i=4,5
\end{equation}
where $c_{i}=\left\{ 2/3,-5/12,1/12,1/2,1/6\right\} $, $d_{4}=1/6$
and $d_{5}=3/2$. The same bag parameters apply for the set of $P_{L}\longleftrightarrow P_{R}$
operators. Remember $P_{L,R}=(1\pm\gamma_{5})/2$. We also use $f_{B_{s}}=230.3\pm1.3\,\mathrm{MeV}$ \cite{FLAG:2021npn} and $M_{B_{s}}=5366.92\pm0.10\,\mathrm{MeV}$ \cite{ParticleDataGroup:2022pth} as input values.

\section{Semileptonic observables}
\label{app:C}
\subsection{\texorpdfstring{$b\to c\tau\nu$}{bcnutau}}
We are interested in the following operators from the effective Lagrangian
describing $b\to c\tau\bar{\nu}$ transition,
\begin{flalign}
\mathcal{L}_{b\rightarrow c\tau\nu} & =-\frac{4G_{F}}{\sqrt{2}}V_{cb}\left[\left(1+\left[C_{\nu edu}^{V,LL}\right]^{3332*}\right)\left[\mathcal{O}_{\nu edu}^{V,LL}\right]^{3332\dagger}\right.\label{eq:btoctaunu_low}\\
 & \left.+\left[C_{\nu edu}^{S,RL}\right]^{3332*}\left[\mathcal{O}_{\nu edu}^{S,RL}\right]^{3332\dagger}+\mathrm{h.c.}\right]\nonumber 
\end{flalign}
where the Wilson coefficients are at the $m_{b}$ scale. The observables
driving the NP effects are the universality ratios $R_{D^{(*)}}$:
\begin{flalign}
R_{D}=R_{D}^{\mathrm{SM}} & \left[\left|1+\left[C_{\nu edu}^{V,LL}\right]^{3332*}\right|^{2}+1.5\mathrm{Re}\left\{ \left(1+\left[C_{\nu edu}^{V,LL}\right]^{3332*}\right)\left[C_{\nu edu}^{S,RL}\right]^{3332*}\right\} \right.\label{eq:RD}\\
 & \left.+1.03\left|\left[C_{\nu edu}^{S,RL}\right]^{3332*}\right|^{2}\right]\,,\nonumber 
\end{flalign}
\begin{flalign}
R_{D^{*}}=R_{D^{*}}^{\mathrm{SM}} & \left[\left|1+\left[C_{\nu edu}^{V,LL}\right]^{3332*}\right|^{2}+0.12\mathrm{Re}\left\{ \left(1+\left[C_{\nu edu}^{V,LL}\right]^{3332*}\right)\left[C_{\nu edu}^{S,RL}\right]^{3332*}\right\} \right.\label{eq:RD*}\\
 & \left.+0.04\left|\left[C_{\nu edu}^{S,RL}\right]^{3332*}\right|^{2}\right]\,,\nonumber 
\end{flalign}
where $R_{D}^{\mathrm{SM}}=0.298\pm0.004$ and $R_{D^{*}}^{\mathrm{SM}}=0.254\pm0.005$
\cite{HFLAV:2022pwe}, and the numerical coefficients are obtained from integrating
over the full kinematical distributions for the $B\rightarrow D^{(*)}$
semileptonic decay \cite{Murgui:2019czp,Mandal:2020htr}. 

\subsection{\texorpdfstring{$b\to s\tau\tau$}{bstautau}\label{sec:b_s_tautau}}

In order to express in a simpler way the observables in the $b\to s\tau\tau$
channel, it is convenient to adopt a different operator basis than the
one in \eqs{eq:basis1}{eq:basis7}. We introduce: 
\begin{align}
\mathcal{O}_{9}^{\tau\tau}= & \left(\bar{s}\gamma_{\mu}P_{L}b\right)\left(\bar{\tau}\gamma^{\mu}\tau\right)\,, & \mathcal{O}_{S}^{\tau\tau}= & \left(\bar{s}P_{R}b\right)\left(\bar{\tau}\tau\right)\\
\mathcal{O}_{10}^{\tau\tau}= & \left(\bar{s}\gamma_{\mu}P_{L}b\right)\left(\bar{\tau}\gamma^{\mu}\gamma_{5}\tau\right)\,, & \mathcal{O}_{P}^{\tau\tau}= & \left(\bar{s}P_{R}b\right)\left(\bar{\tau}\gamma_{5}\tau\right)\,,
\end{align}
which give rise to the following effective Lagrangian: 
\begin{equation}
\begin{aligned}\mathcal{L}_{b\rightarrow s\tau\tau}=\frac{4G_{F}}{\sqrt{2}}V_{ts}^{*}V_{tb}\frac{\alpha}{4\pi} & \bigg[(C_{9}^{\mathrm{SM}}+C_{9}^{\tau\tau})\mathcal{O}_{9}^{\tau\tau}+(C_{10}^{\mathrm{SM}}+C_{10}^{\tau\tau})\mathcal{O}_{10}^{\tau\tau}\\
 & +C_{S}^{\tau\tau}\mathcal{O}_{S}^{\tau\tau}+C_{P}^{\tau\tau}\mathcal{O}_{P}^{\tau\tau}+\mathrm{h.c.}\bigg]\,,
\end{aligned}
\end{equation}
where for simplicity we suppressed the scale dependence of the Wilson
coefficients, which are at the scale $\mu=m_{b}$. The SM values are
given by $C_{9}^{\mathrm{SM}}=4.27$ and $C_{10}^{\mathrm{SM}}=-4.17$ \cite{Bruggisser:2021duo}. With these definitions, the
expressions for the observables of interest read \cite{Becirevic:2016zri,Cornella:2021sby}
\begin{flalign}
\mathcal{B}(B_{s}\rightarrow\tau\tau) & =\mathcal{B}(B_{s}\rightarrow\tau\tau)_{\mathrm{SM}}\bigg\{\left|1+\frac{C_{10}^{\tau\tau}}{C_{10}^{\mathrm{SM}}}+\frac{C_{P}^{\tau\tau}}{C_{10}^{\mathrm{SM}}}\frac{M_{B_{s}}^{2}}{2m_{\tau}\left(m_{b}+m_{s}\right)}\right|^{2}\nonumber \\
 & +\left(1-\frac{4m_{\tau}^{2}}{M_{B_{s}}^{2}}\right)\left|\frac{C_{S}^{\tau\tau}}{C_{10}^{\mathrm{SM}}}\frac{M_{B_{s}}^{2}}{2m_{\tau}\left(m_{b}+m_{s}\right)}\right|^{2}\bigg\}\label{eq:Bs_tautau}\\
\mathcal{B}(B^{+}\rightarrow K^{+}\tau\tau) & =10^{-9}\bigg(2.2\left|C_{9}^{\tau\tau}+C_{9}^{\mathrm{SM}}\right|^{2}+6.0\left|C_{10}^{\tau\tau}+C_{10}^{\mathrm{SM}}\right|^{2}+8.3\left|C_{S}^{\tau\tau}\right|^{2}\nonumber \\
 & +8.9\left|C_{P}^{\tau\tau}\right|^{2}+4.8\mathrm{Re}[C_{S}^{\tau\tau}(C_{9}^{\tau\tau}+C_{9}^{\mathrm{SM}})^{*}]+5.9\mathrm{Re}[C_{P}^{\tau\tau}(C_{10}^{\tau\tau}+C_{10}^{\mathrm{SM}})^{*}]\bigg)\,,
\end{flalign}
where we use $\mathcal{B}(B_{s}\rightarrow\tau\tau)_{\mathrm{SM}}=\left(7.73\pm0.49\right)\cdot10^{-7}$
\cite{Bobeth:2013uxa} and $\mathcal{B}(B^{+}\rightarrow K^{+}\tau\tau)_{\mathrm{SM}}=\left(1.4\pm0.2\right)\cdot10^{-7}$
\cite{Bouchard:2013mia}. The numerical values for the NP contributions
to $B^{+}\to K^{+}\tau\tau$ decays are taken from \cite{Bouchard:2013mia,Cornella:2021sby}.

The operators $C^{\tau\tau}_{9}$ and $C^{\tau\tau}_{10}$ are related to
$[C_{ed}^{V,LL}]^{3323}$ and $[C_{de}^{V,LR}]^{3323}$ via

\begin{align}
C_{9}^{\tau\tau}=\frac{2\pi}{\alpha V_{ts}^{*}V_{tb}}\left([C_{ed}^{V,LL}]^{3323}+[C_{de}^{V,LR}]^{3323}\right)\,, &  & C_{10}^{\tau\tau}=\frac{2\pi}{\alpha V_{ts}^{*}V_{tb}}\left([C_{de}^{V,LR}]^{3323}-[C_{ed}^{V,LL}]^{3323}\right)\,. \label{eq:C9C10_CedV}
\end{align}

The operators $[C_{ed}^{S,LL}]^{3323}$ and $[C_{ed}^{S,LR}]^{3323}$ are related to $C_{S}^{\tau\tau}$
and $C_{P}^{\tau\tau}$ via

\begin{align}
C_{S}^{\tau\tau}=\frac{2\pi}{\alpha V_{ts}^{*}V_{tb}}\left([C_{ed}^{S,RR}]^{3323}-[C_{ed}^{S,LR}]^{3323}\right) &  & C_{P}^{\tau\tau}=\frac{2\pi}{\alpha V_{ts}^{*}V_{tb}}\left([C_{ed}^{S,RR}]^{3323}+[C_{ed}^{S,LR}]^{3323}\right)  \label{eq:CSCP_CedS}
\end{align}

\bibliographystyle{utphys}
\bibliography{references}

\end{document}